\documentclass[aps,pre,twocolumn,showpacs,final,floatfix]{revtex4-1}

\usepackage[pdftex]{graphicx}  
\usepackage[update]{epstopdf}
\usepackage{amssymb}
\usepackage{amsmath}
\usepackage{color}
\usepackage{textcomp}
\usepackage[utf8]{inputenc}
\newcommand\f{\frac}
\newcommand\ra{\rangle}
\newcommand\la{\langle}
\newcommand\nn{\nonumber}

\newcommand{\eqa}[1]{\begin{align}#1\end{align}}
\graphicspath {{./figures/}}
\makeatletter
\def\input@path{{./figures/}}
\makeatother

\begin{document}
 \title{Revisiting the  Mazur bound and the Suzuki equality}

\author{Abhishek Dhar}
\affiliation{International Centre for Theoretical Sciences, Tata Institute of Fundamental Research, Bengaluru -- 560089, India}
\author{Aritra Kundu}
\affiliation{SISSA -- International School for Advanced Studies and INFN, via 
	Bonomea 265, 34135 Trieste, Italy}
\author{Keiji Saito}
\affiliation{Department of Physics, Keio University, Yokohama 223-8522, Japan} 

\date{\today}

\begin{abstract}
 Among the few known rigorous results for time-dependent equilibrium correlations, important for understanding transport properties, are the Mazur bound and the Suzuki equality.   
The Mazur inequality gives a lower bound, on the long-time average of  the time-dependent auto-correlation function of observables, in terms of equilibrium correlation functions involving conserved quantities. On the other hand, Suzuki proposes an exact equality for quantum systems. In this paper, we discuss the relation between the two results and in particular, look for the analogue of the Suzuki result for classical systems. This requires us to examine as to what constitutes a complete set of conserved quantities required to saturate the Mazur bound. We present analytic arguments as well as illustrative numerical results from a number of  different systems. Our examples include systems with few degrees of freedom as well as many-particle integrable models, both free and interacting.
\end{abstract}

\maketitle

 \section{Introduction}
Time dependent equilibrium auto-correlation functions of physical observables play an important role in understanding dynamical properties of a system. In particular they tell us about the ergodicity of a  given Hamiltonian system. In a seminal paper \cite{mazur1969}, Mazur discussed the long time average of such auto-correlation functions in the context of ergodicity.  A rigorous lower bound was obtained for this quantity. In the classical description, we consider systems with phase space degrees of freedom $(x,p)=\{x_i,p_i\},~i=1,2,\ldots,N$ and $A=A(x,p)$ is some function of the phase space variables and describes some physical observable. The systems  dynamics is described by a Hamiltonian $H(x,p)$ and let us assume that apart from $H$ there are $R-1$ other conserved quantities, \emph{not necessarily independent ones}. We denote the set of conserved quantities as   ${\bf I}_R = (I_1,I_2,\ldots,I_R)$, with $I_1=H - \la H \ra$, where $\la \ldots \ra$ denotes a thermal average over the Gibbs distribution $\rho=e^{-\beta H}/Z$ with $Z= \int dx dp e^{-\beta H}$. Without loss of generality, one can assume that $\la I_k \ra=0$ for all $k$. We define the correlation matrix, $C$, with elements $C_{ij}=\la I_i I_j \ra$, 
and consider the quantity
\begin{align}
D_A := \sum_{k,\ell} \langle I_k A \ra [C^{-1}]_{k \ell} \la I_\ell A \ra,  \label{MB}
\end{align}
which we will refer to as the Mazur bound. 
Let us also define the long time average of the temporal auto-correlation function of the observable $A$:
\begin{align}
C_A := \lim_{\tau \to \infty} \f{1}{\tau} \int_0^\tau dt \la A(t) A(0) \ra,  
\label{LTA}
\end{align}
where the average is again over the equilibrium Gibbs distribution.
 Mazur proved that
\begin{align}
C_A \geq D_A.
\label{mazur}
\end{align}

In \cite{mazur1969} it was shown that this result could provide insights on the ergodicity of the variable or its absence. We briefly discuss the notion of ergodicity as indicated in the behavior of the correlation function. Let $\bar{A}(E)=\int dq dp~ A(q,p)~ \delta (E-H)/\int dq dp ~ \delta (E-H)$ denote the microcanonical average of the observable. Then ergodicity implies $C_A= \la \bar{A}^2(E) \ra$ and in the thermodynamic limit this leads to the equality,
\begin{align}
C_A = \f{\la A \Delta H \ra^2}{\la (\Delta H)^2 \ra}, \label{ergodic}
\end{align}
where $\Delta H= H -\la H \ra$. 
Allowing for the presence of extra conservation laws, the notion of sub-ergodicity was discussed in \cite{vankampen}, which in modern terms relates to the idea of generalized Gibbs ensembles.

The Mazur bound can be proven for both classical and quantum systems as already noted in the original paper. For the case of quantum systems, an exact Mazur-type equality was derived by Suzuki, which hold when one includes a ``sufficient'' number of constants of motion. For example, in a quantum system with a  Hilbert space of finite dimension $\mathcal{D}$, a trivial complete set of constants of motion are the energy projection operators $\hat{P}_n=|n\ra \la n |$ where $|n\ra$, with $n=1,2,\ldots,\mathcal{D}$, denotes the energy eigenstates.  
Then (see later) it is easy to show that one obtains the equality in the Mazur relation in Eq.~\eqref{mazur}. Two natural questions that one could ask are: (i) instead of the projection operators, is it possible to obtain the equality with a smaller number of ``local'' constants of motion and (ii) does this result have a classical analogue. One of the aims of the present work is to discuss these questions and provide illustrative examples that clarify some subtle issues.

An important application of the Mazur relation has been in the context of transport properties of integrable systems \cite{zotos1997,zotos2002,affleck2011,prosen2013,nardis2017,doyon2017}. The auto-correlation functions involving currents corresponding to conserved quantities  are related to transport coefficients via the Green-Kubo formulas. In particular,  the asymptotic long time saturation value of the auto-correlation, $C_A$, gives the so-called  Drude weight  which is the strength of the zero-frequency component of the conductivity and implies ballistic transport. It was pointed out in  \cite{zotos1997} that the Mazur bound can be used to prove the presence of a finite Drude weight for integrable systems.  This was used to prove ballistic transport in one-dimensional systems such as the quantum spin-$1/2$ XXZ chain \cite{prosen2011} and the Toda lattice \cite{zotos2002}. 
For classical integrable systems with $N$ degrees of freedom, the set of exactly $N$ \emph{independent} constants of motion, which we denote by $\{Q_i\}$, is in some sense special and in fact their existence defines integrability (e.g one can construct action-angle variables). One might expect that this set should be sufficient to saturate the Mazur bound. However, the numerical study in \cite{young2010} found that one needs to include bilinear combinations of the conserved  quantities, of the form  $Q_j Q_k$, in order to approach the equality. The study was restricted to systems of sizes $N=4,6,8$ and an important question is whether the contribution of the bilinear terms vanishes in the thermodynamic limit.
On the other hand, the situation is even more complicated in the quantum case since the notion of quantum integrability is not so well defined and a basic question is on the choice of the set of constants ${\bf I}$ required to saturate the Mazur bound. These aspects will also be discussed in this paper. 

The plan of the paper is as follows: In Sec.~\eqref{sec:proofs} we outline the proofs of the Mazur inequality and the Suzuki equality. In Sec.~\eqref{sec:analogue} we describe a procedure which leads to the classical analogue  of the Suzuki equality. As illustrative examples, we then provide in Sec.~\eqref{sec:examples} explicit results, both  numerical and analytical, on the application of the Mazur-Suzuki results in different  physical systems. These include few body systems such as an oscillator and a two coupled spin system, as well as many body systems such as systems described by quadratic Hamiltonians, and finally the Toda chain.  We conclude with a discussion in Sec.~\eqref{sec:summary}.

\section{Proof of the Mazur and Suzuki relations}
\label{sec:proofs}
{\bf The Mazur bound}: We start the discussion for a classical system. Consider an observable $Y$ whose time evolution is given by $Y(t)=Y(q_t,p_t)$ and let $\langle ...\rangle$  denotes an average over the canonical distribution $e^{-\beta H}/Z$. Using stationarity $\la Y(t_1) Y(t_2)\ra=\la Y(t_1-t_2) Y(0)\ra$ and time reversal invariance $\la Y(t) Y(0) \ra = \la Y(-t) Y(0)\ra$ one can show, via the Wiener-Khinchine theorem, that the correlation can be expressed in terms of the power spectral density of the signal. Thus one has 
\begin{align}
\la Y(t) Y(0)\ra &= \int_{-\infty}^\infty df S(f) e^{-2 \pi i f t} \\
{\rm where}~~S(f)&= \lim_{\tau \to \infty} \f{1}{\tau} \left| \int_{-\tau/2}^{\tau/2} Y(t) e^{-2 \pi i f t} \right|^2 \geq 0
\end{align}
Application of the Wiener-Khinchine theorem then leads to the result 
\begin{align}
\overline{\langle Y(t) Y(0)\rangle}= \lim_{\tau \to \infty} \frac{1}{\tau}\int_0^\tau \langle Y(t) Y(0) \rangle= S(0) \ge 0~.\label{WK-C}
\end{align}
For a quantum system where  now $Y,H$ are  now  Hermitian operators and with time evolution of $Y$ given by $Y(t)=e^{i H t} Y e^{-iHt}$ we note that
\begin{align}
\langle Y(t) Y(0) \rangle = \sum_{n,m} |Y_{n,m}|^2 e^{i(E_n-E_m)t/\hbar} \frac{e^{-\beta E_n}}{Z},
\end{align}
where $Y_{n,m}=\langle n |Y|m\rangle$ are the matrix elements of the operator in the energy basis specified by states $|n\rangle$ and eigenvalues $E_n$. 
Performing a time average, all oscillatory terms with $E_n \neq E_m$ vanish and we get
\begin{align}
\lim_{\tau \to \infty} \frac{1}{\tau}\int_0^\tau \langle Y(t) Y(0)
= \sum_{n,m}^{E_n=E_m} |Y_{n,m}|^2 \frac{e^{-\beta E_n}}{Z} \rangle \ge 0~.\label{WK-Q}
\end{align}

Let us consider that our system has a set of  conserved quantities $I_k~,k=1,2,\ldots,R$ satisfying $\la I_k \ra=0$, that we denote as ${\bf I}_R$ and  define the correlation matrix 
\begin{equation}
C_{ij}=\langle I_i I_j \rangle~.
\end{equation}
The correlation matrix is positive definite, so has a positive determinant and is invertible. In this case  Mazur proves that for an observable $A$, one has the following bound:
\begin{align}\label{eq:mazur}
C_A=\overline{\la A(t) A(0)\ra} \geq \sum_{k,\ell} \langle I_k A \ra [C^{-1}]_{k \ell} \la I_\ell A \ra~. 
\end{align}
We denote the quantity on the right hand side, constructed out of $R$ conserved quantities, as $M^{A}_R$.  
The proof, valid for both classical and quantum systems, starts with the inequality in Eqs.~(\ref{WK-C},\ref{WK-Q}). Let us take $Y=A-\sum_{k=1}^R z_k I_k= A- z^T I$, where we denote any set of $R$ constants $\{ I_k\}$ by the column vector $I$ and $z^T=(z_1,z_2,\ldots,z_R)$ is a set of arbitrary real numbers. Then we get 
\begin{align}
\overline{\langle Y(t) Y(0)\rangle} &= \overline{\langle (A(t)-z^TI)(A(0)-I^Tz) \rangle} \nn \\
&= \overline{\langle A(t)A(0) \rangle} +z^T \langle I I^T \rangle z -2 \langle A I^T\rangle z~.
\end{align} 
This quadratic form, in the variables $z$, is minimized  for the choice $z=C^{-1} \langle A I\rangle$  which gives 
\begin{align}
\overline{\langle Y(t) Y(0)\rangle}=\overline{\langle A(t)A(0) \rangle}- \langle A I^T\rangle C^{-1} \langle A I\rangle~.
\end{align}
Using Eq.~(\ref{WK-C}) or Eq.~\eqref{WK-Q} we then immediately get the Mazur inequality Eq.~(\ref{eq:mazur}).
Note that, without loss of generality, we can consider a new  set of constants labeled ${J}_k$, linearly related to the earlier set  $J=  O^T I $ by the orthogonal transformation $O$  which diagonalizes the correlation matrix $C$. Then we get $\langle J J^T \rangle = O^T \langle I I^T \rangle O = O^T C O= Diag[\langle J_k^2\rangle]$. And then $\langle A I^T \rangle C^{-1} = \langle A I\rangle = 
\langle A J^T \rangle O^T C^{-1} O \langle A J \rangle = \sum_k  { \langle J_k A\rangle^2}/{\langle J_k^2 \ra}$.  
Hence, without loss of generality we can write Eq.~\ref{eq:mazur} in the form
\begin{align}
C_A \geq \sum_k \frac{\langle J_k A\rangle^2}{\langle J_k^2 \rangle}~. 
\end{align}

{\bf The Suzuki equality}: On the other hand, Suzuki considers quantum systems with a discrete energy spectrum with energy eigenstates labeled as $|\alpha \rangle$ and eigenvalues $\epsilon_\alpha$ with $\alpha=1,2,\ldots$, and the Hilbert space could be finite or infinite dimensional.  Here we restrict ourselves to the case of systems of finite dimensions (see \cite{prosen2013} for a discussion of the more general case). Then let us assume that there exist constants of motion $I_k~,k=1,2,\ldots,R$  such that we can decompose the operator as
\begin{align}
\hat{A}= \sum_k c_k \hat{I}_k + \hat{A}'~,
\end{align}
where, in the energy eigenbasis,  the  operator $\hat{A}'$ satisfies $\langle \alpha| \hat{A}' |\alpha \rangle =0$ for all $\alpha$ and  $\langle \alpha| \hat{A}' |\alpha'\rangle =0$ for degenerate levels with $\epsilon_\alpha = \epsilon_\alpha'$. Then using $Y=A$ in Eq.~\eqref{WK-Q}, Suzuki proves the \emph{equality}
\begin{align}
C_A = \sum_{k,\ell} \la I_k A \ra [C^{-1}]_{k \ell} \la I_\ell A \ra = \sum_k \frac{\langle J_k A\rangle^2}{\langle J_k^2 \rangle}~,  \label{eq:suzEq}
\end{align}
where the second equality again follows on choosing the set ${\bf J}$ as  linear combinations of the set ${\bf I}$ so that the correlation matrix $C$ is diagonal.
A trivial choice of the set $J_k$ is simply to choose them as the projection operators corresponding to energy eigenstates (which is chosen to be orthonormal), i.e, we choose 
\begin{align}
J_n= |n><n|~, 
\end{align}
where $|n \rangle$ runs through the full set of $\mathcal{D}$ energy eigenstates. Note that even when degeneracies are present, we can always choose a linear combination of the degenerate states such that the conditions on $A'$ are satisfied. To get the strict equality, we need to take the full set of eigenstates. Thus, for systems with an infinite Hilbert space,  such as a harmonic oscillator we need to consider(for general operators $A$)  an infinite number of conserved quantities.   

As a special case consider a quantum system with a finite dimensional Hilbert space of dimensions $\mathcal{D}$. Let us assume that the Hamiltonian is the only conserved quantity and the eigenvalues are non-degenerate and we have a complete basis of projection operators $J_\alpha =   |\alpha><\alpha|$, with $\alpha=1,2,\ldots \mathcal{D}$. Then we have $\sum_\alpha J_\alpha = I,~ \sum_\alpha \epsilon_\alpha J_\alpha = H,  \sum_\alpha \epsilon_\alpha^2 J_\alpha = H^2,$ etc. and so we can write:
\begin{align}
\left(
\begin{array}{c}
I \\ H \\H^2 \\. \\. \\. \\ H^{\mathcal{D}-1}
\end{array}
\right) =
\left(
\begin{array}{ccccccc}
1 & 1& 1& . &.&.& 1 \\
\epsilon_1 &\epsilon_2 & \epsilon_3& . &.&.& \epsilon_{\mathcal{D}} \\
\epsilon_1^2 &\epsilon_2^2 & \epsilon_3^2& . &.&.& \epsilon_{\mathcal{D}}^2 \\
. & . & . & . &.&.& . \\
. & . & . & . &.&.& . \\
. & . & . & . &.&.& . \\
\epsilon_1^{\mathcal{D}-1} &\epsilon_2^{\mathcal{D}-1}  & \epsilon_3^{\mathcal{D}-1} & . &.&.& \epsilon_{\mathcal{D}}^{\mathcal{D}-1} 
\end{array}
\right)
\left(
\begin{array}{c}
J_1 \\ J_2 \\J_3 \\. \\. \\. \\ J_{\mathcal{D}}
\end{array}
\right). \label{Hbasis}
\end{align}
For a non-degenerate spectrum, the determinant of the matrix above is non-vanishing and so we can invert the above equation to express the $J_\alpha$s in terms of $H$ and its higher powers. This means that for a generic quantum system with $H$ as the only conserved quantity, the choice ${\bf I}=(I,H,H^2,\ldots,H^{\mathcal{D}-1})$ will provide an equality for the corresponding Mazur bound.  

Now we consider the following two questions, which are closely related:\\
(a) Is there a classical analogue to the Suzuki equality Eq.~(\ref{eq:suzEq})? \\
(b) In the Suzuki equality, is it possible to replace the projection operators by more conventional conserved quantities, for e.g involving local operators?  We expect that a finite and smaller number of such observables can give stronger bounds or exact equality than the energy projectors. 

We shall attempt to answer the first question in the next section and then, in Sec.~\eqref{sec:examples}, we will discuss specific examples which throw some light on both these questions.

\section{Classical analogue of Suzuki equality}
\label{sec:analogue}

Consider a classical Hamiltonian system with $N$ positional and $N$ momentum degrees of freedom, and having $r$ \emph{independent} conserved quantities $\{Q_j\}$. Then the infinite time average
\begin{equation}
\bar{A}(x,p) = \lim_{\tau\to \infty} \frac{1}{\tau} \int_0^\tau dt A(x_t,p_t)~,
\end{equation}
where $(x,p)$ is the initial condition,  is by definition a conserved quantity. 
Let us also define the average of $A$ in a  ``generalized'' microcanonical ensemble as
\begin{align}
\la A \ra_m({\bf q}) = \f{\int dx dp A(x,p) \prod_{k=1}^r \delta(Q_k(x,p)-q_k)}
{\int dx dp \prod_{k=1}^r \delta(Q_k(x,p)-q_k)},
\end{align}
where $q_k$, $k=1,2,\ldots,r$,  are the constrained values of the constants of motion. For an ergodic function the time average, obtained by starting from almost any initial condition satisfying the constrained values of the constants of motion, should be equal to the microcanonical average, i.e
\begin{align}
\bar{A}(x,p)=\la A \ra_m ({\bf Q})= a^{(0)}+\sum_k a^{(1)}_k Q_k + \sum_{k,l}  a^{(2)}_{kl} Q_k Q_l +\ldots~, 
\end{align} 
where we assume that a Taylor series expansion of $\la A\ra_m({\bf I}$ is possible. This then implies that we can write
\begin{align}
A(x_t,p_t)&= a^{(0)}+\sum_k a^{(1)}_k Q_k + \sum_{k,l}  a^{(2)}_{kl} Q_k Q_l +\ldots \nn \\ &+ g(x_t,p_t), 
\end{align} 
where $g(t)=g(x_t,p_t)$ includes oscillatory contributions which average to zero. 
Constructing the set of conserved quantities  $\{I_s\}=(1, Q_1,Q_2,...Q_r,Q_1^2, Q_1 Q_2,..., Q_1^3,...)$, we thus see that, in general we require the following infinite series expansion
\begin{align}
\bar{A}(x,p)=\sum_s c_s I_s~. \label{eq:bAexp}
\end{align}
We now define averages $\la ...\ra$ over the generalized Gibbs ensemble defined by $\rho(x,p)= e^{-\sum_{k=1}^r \lambda_k Q_k}/Z_{GGE}$, where $\lambda_k$ are  intensive parameters that are conjugate to the variables $Q_k$.  Clearly this is an invariant measure and can be obtained from a corresponding generalized microcanonical ensemble. Taking averages over this distribution, we get
$c_r =\sum_{s} [C^{-1}]_{rs} \la \bar{A} I_s \ra$ 
where $C_{rs}=\la I_r I_s\ra$.
We further observe that
\begin{align}
\la \bar{A} I_r \ra&= \lim_{\tau \to \infty} \frac{1}{\tau} \int_0^\tau dt \la A(t) I_r \ra \nn \\
&=\lim_{\tau\to \infty} \frac{1}{\tau} \int_0^\tau dt \la A(t) I_r(t) \ra = \la A I_r \ra~,
\end{align}
where we used  $I_r=I_r(t)$  since this is a constant of motion, and the fact that the averaging is over a time invariant distribution.  Thus, in matrix form, we have $c=C^{-1} \la A I\ra$. Using this and Eq.~\eqref{eq:bAexp} we finally get
\begin{align}
&\lim_{ \tau \to \infty} \frac{1}{\tau} \int_0^\tau dt \la A(t) A(0) \ra \nn \\ 
&= \int \prod_{k=1}^r d Q_k~e^{-\sum_k \lambda_k Q_k}  \lim_{ \tau \to \infty} \frac{1}{\tau} \int_0^\tau dt  \la A(t) A(0) \ra_m({\bf Q}) \nn \\ 
&~~~=\int \prod_{k=1}^r d Q_k~e^{-\sum_k \lambda_k Q_k}  \bar{A}^2 
= \la \bar{A}^2 \ra  \nn \\
&=\sum_{r,s} c_r c_s C_{rs}= c^T C c= \la A I^T\ra C^{-1}  \la A I\ra~. \label{mazur-suzuki}
\end{align}
This then is the expected form of the Suzuki equality for a classical system, the main point being that it is not sufficient to take a finite number of conserved quantities but, in general, it is required to take an infinite set comprising of powers of the standard independent conserved quantities. Secondly, we used the notion of ergodicity within the generalized constant ${\bf Q}$ ensemble and this is a necessary condition for the equality to be obtained. Note that while the above result has been derived for the GGE, the special case with $Q_1=H, \lambda_1=\beta$ and $\lambda_k=0$ for $k=2,3,\ldots,R$, recovers the case with the  usual Gibbs ensemble. 

In the next section we will discuss specific examples to illustrate these points.

We note that for a classical system the usual definition of integrability for a system of $2 N$ degrees of freedom
is that there are $N$ ``independent'' conserved quantities (i.e with vanishing Poisson brackets). This ensures that the system has a description in terms of action-angle variables and the Lyapunov exponent vanishes. However, the notion of ``independent'' conserved variables is different as far as the Mazur relation is concerned. Independence is now  defined in terms of the scalar product $\la I_k I_\ell \ra$. We show below that we can add  a new conserved variable  which is orthogonal to the existing set and this will always improve the bound, provided  this new variable has some overlap with the measured observable $A$.    

{\bf Results on bounds}:
Let us denote the bound obtained for the choice of the orthonormal set  $(J_1,J_2,\ldots,J_{n-1})$ by $B_{n-1}$. We add another conserved quantity $I_n$ and ask as to how the new bound $B_n$ changes. 

\emph{Proof that $B_n > B_{n-1}$}:
Note that the new vector $I_n$ need not be orthogonal to the earlier vectors, so that in general $\langle I_n J_i \rangle \neq 0$ for $i=1,2,\ldots,n-1$. Let $\Delta_n$ be the determinant of the correlation matrix $C$. Then one can show
\begin{align}
(B_n-B_{n-1})\Delta_n=\left[ \la A I_n \ra  - \sum_{k\neq n} \la A J_k\ra  \frac{\la I_n J_k \ra}{\la J_k^2 \ra}\right]^2\prod_{k=1}^{n-1}\la J_k^2\ra ~. \label{eq:conv}
\end{align}
This proves that the addition of  any linearly independent vector will in general improve the bound. We can construct the new vector $I_n=J_n$ such that it is orthogonal to the previous existing set. In that case Eq.~(\ref{eq:conv}) leads to the expected result
\begin{align}
B_n-B_{n+1}= \frac{\la A J_n\ra^2}{\la J_n^2 \ra}
\end{align}   

\section{Examples}
\label{sec:examples}
In this section we discuss a number of examples to illustrate and clarify the Mazur-Suzuki bounds and their applications. 

\subsection{Classical anharmonic oscillator with a single conserved quantity}
We consider a single anharmonic oscillator described by the Hamiltonian
\begin{align}
H=\frac{p^2}{2}+k\frac{x^2}{2}+\alpha \frac{x^4}{4}~. \label{eq:Hanosc}
\end{align}
In this case we note that the system is always ergodic even when  the non-linear
term is absent ($\alpha=0$).

Let us first consider the harmonic case with $\alpha=0$, for which both $C_A$ and $D_A$ can be computed exactly. In this case, if we set $A=x^2$ then $C_A= 2\la x^2 \ra^4$ and we can also verify that $D_A=\la A H\ra^2/\la H^2 \ra = 2 \la x^2 \ra^4$ so we see that the Mazur  equality is satisfied with the choice   ${\bf I}= \{H\}$. However, for the observable $A=x^4$ we find $C_A=54 \la x^2\ra^2$ but $\la A H\ra^2/\la H^2 \ra = (81/2) \la x^2 \ra^2$. In this case, one can easily verify that the Mazur equality occurs for the choice ${\bf I}=  \{H, H^2 \}$.

\begin{figure}
  \centering
  \includegraphics[width=0.9\columnwidth]{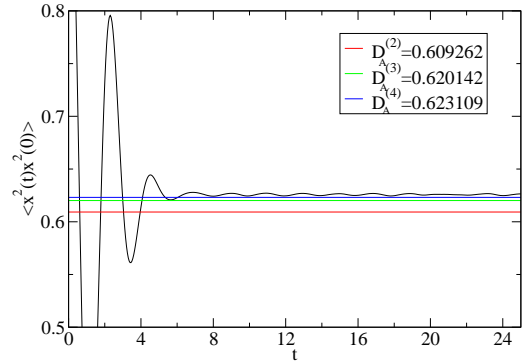}
  \caption{Classical anharmonic oscillator: Parameter values --- $k = 0.0, \alpha=1.0$. Plots of auto-correlation functions for  of the observable $A=x^2$, as obtained from direct simulations, compared with Mazur bounds for different choices of sets of conserved variables: $D_A^{(2)}=0.609262$ is for the set ${\bf I}_2=\{1,H \}$, while   $D_A^{(3)}=0.620142$ is for the set ${\bf I}_3=\{1,H,H^2\}$, $D_A^{(4)}=0.623109$ is for the set ${\bf I}_4=\{1,H,H^2, H^3 \}$ . We see a rapid convergence to $C_A$, the saturation value of the autocorrelation $A(t)A(0)$.}
  \label{figanh}
\end{figure}

For the anharmonic case with $k=0$ and $\alpha \neq 0$, it is no longer possible to compute either $C_A$ or $D_A$ analytically. For $A=x^2$ we plot in Fig.~\eqref{figanh} the results from simulations and compare with the Mazur bound for different choices of the set ${\bf I}$.  Performing the required integrations numerically we find the   following Mazur bounds for different ${\bf I}$: $D_A=0.456947$ for the set ${\bf I}_1=\{ 1 \}$, $D_A^{(1)}=0.543984$ for the set  ${\bf I}_1=\{ H \}$, $D_A^{(2)}=0.609262$ is for the set ${\bf I}_2=\{1,H \}$, while   $D_A^{(3)}=0.620142$ for the set ${\bf I}_3=\{1,H,H^2\}$, $D_A^{(4)}=0.623109$ for the set ${\bf I}_4=\{1,H,H^2, H^3\}$, and $D_A^{(5)}=0.624347$ for  the set ${\bf I}_5=\{1,H,H^2,H^3,H^4\}$. With increasing number of the conserved charges, we see a clear convergence of the  Mazur bound to the numerically obtained saturation value for the autocorrelation, $C_A$. 

{\bf Main conclusions}: For the harmonic case, a finite number of conserved quantities in the set ${\bf I}$ is sufficient to saturate the Mazur bound. However, the number of conserved quantities required, depends on the degree of the observable. In the nonlinear case, for any observable, one requires an infinite number of conserved variables, though   the convergence to the Mazur bound is quite fast.

\subsection{Two site $XXZ$ model: classical and quantum}

\begin{figure}
  \centering
  \includegraphics[width=0.9\columnwidth]{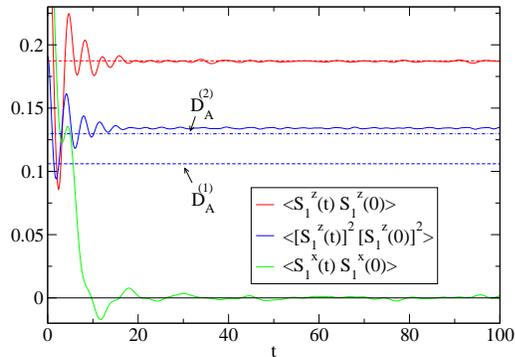}
  \caption{Classical spins: Parameter values --- $\Delta = 0.5, h=0.0$. Plots of auto-correlation functions for three different choices of the observable $A$, as obtained from direct simulations, compared with Mazur bounds for different choices of sets of conserved variables.  The  Mazur bound $D_A^{(1)}$  is for the set ${\bf I}=\{1,S^z, H\}$, while   $D_A^{(2)}$ is for the set ${\bf I}=\{1,S^z,H,(S^z)^2,S^zH,H^2\}$.  For the observables $A=S_1^z$, $D_A^{(1)}=D_A^{2)}=0.187368$, while for $A=S_1^x$ we get $D_A^{(1)}=D_A^{2)}=0$, and the horizontal lines in the plot indicate these values. On the other hand, for the observable $A=[S_1^{(z)}]^2$ we get $D_A^{(1)}=0.106008$ $D_A^{2)}=0.129726$.}
  \label{fig1}
\end{figure}

\begin{figure}
  \centering
  \includegraphics[width=0.9\columnwidth]{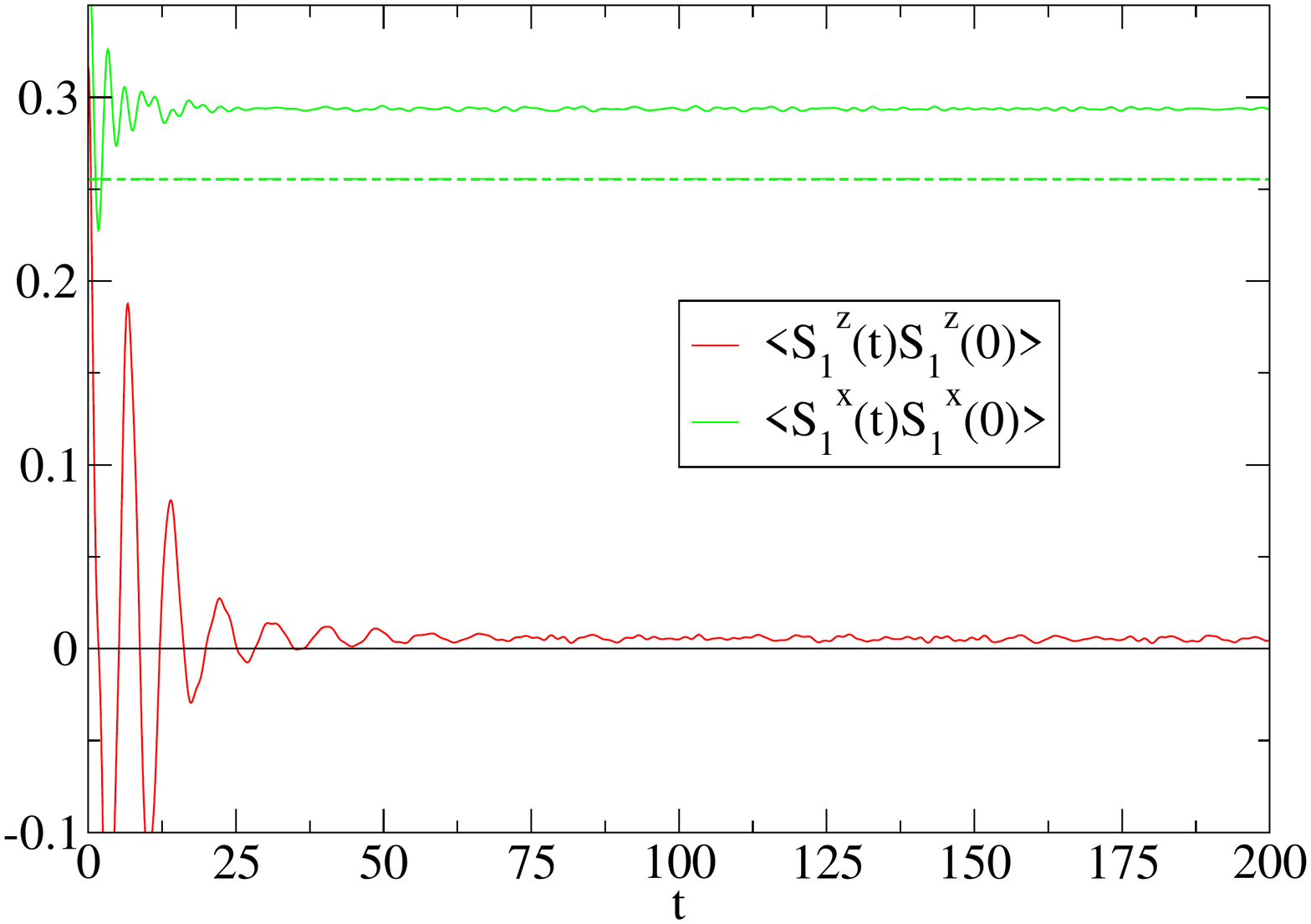}
  \caption{Classical spins: Parameter values --- $\Delta = 1.5, h=1.0$. Plots of auto-correlation functions for three different choices of the observable $A$, as obtained from direct simulations, compared with Mazur bounds for different choices of sets of conserved variables.  The  Mazur bound $D_A^{(1)}$  is for the set ${\bf I}=\{1, H\}$,  $D_A^{(2)}$ is for the set ${\bf I}=\{1,H,H^2\}$ and  $D_A^{(3)}$ is for the set ${\bf I}=\{1,H,H^2,H^3\}$ . We get $D_A^{(1)}=D_A^{(2)}=D_A^{(2)}=0$ for $A=S_1^z$,
$D_A^{(1)}=0.255225$, $D_A^{(2)}=0.2556429,D_A^{(3)}=0.25564436$ for $A=S_1^x$.  We see that the saturation values are close to the bounds (as we include higher order terms) but we do not get convergence on including more terms.}
  \label{fig2}
\end{figure}

\begin{figure}
  \centering
  \includegraphics[width=0.9\columnwidth]{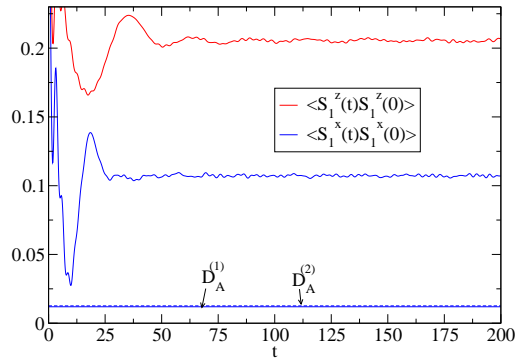}
  \caption{Classical spins: Parameter values --- $\Delta = 1.5, h=0.1$. Plots of auto-correlation functions for two different choices of the observable $A$, as obtained from direct simulations, compared with Mazur bounds for different choices of sets of conserved variables.  The  Mazur bounds    $D_A^{(1)}$ is for the set ${\bf I}=\{1,H,H^2\}$, while  $D_A^{(2)}$ is for the set ${\bf I}=\{1,H,H^2,H^3\}$ . We get $D_A^{(n)}=0$ for $A=S_1^z$ at all orders, while for   $A=S_1^x$ we get $D_A^{(1)}=0.0119538$ and  $D_A^{(2)}=0.0126232$. Thus we see that the saturation values are far from the bounds and the system is highly non-ergodic.}
  \label{fig3}
\end{figure}

We consider the $XXZ$ spin model with a transverse external magnetic field described by the following Hamiltonian
\begin{align}
H=-(S^x_1 S^x_2+S^y_1 S^y_2) -\Delta (S^z_1 S^z_2) -h (S^x_1+S^x_2)~.\label{eq:Hspin} 
\end{align}
For $h=0$ the system has two conserved quantities $H$ and $S^z=S^z_1+S^z_2$, while for $h \neq 0$, we expect $H$ to be the only conserved quantity. As the physical observable we consider the following three quantities:
\begin{align}
A=S^z_1,~(S_1^z)^2,~S_1^x~. 
\end{align}

{\bf Classical case}: 
In Figs.~(\ref{fig1},\ref{fig2},\ref{fig3}) we show plots of the corresponding correlation functions for the parameter sets  $(\Delta ,h)=(0.5,0),~(1.5,1.0),~(1.5,0.1)$. For the integrable case, we use the set ${\bf I}=(1,S^z,H,(S^z)^2,S^zH,H^2,...)$ as our basis set for the Mazur bound, while for the non-integrable case, when $h\neq0$, we use the set ${\bf I}=(1,H,H^2,...)$. We set $\beta=1$ in all cases. We find that for the integrable case $h=0$, shown in Fig.~\eqref{fig1}, we get convergence to the Mazur bound with the basic set of independent conserved quantities ${\bf I}=(1,S^z,H)$, for the case of the observables $S_1^z$ and $S_1^x$. On the other hand, for the observable $[S_1^z]^2$, convergence to the Mazur value requires us to increase the number of conserved variables by considering products. Thus, this system exhibits ergodicity within the restricted phase space of constant $H$ and $S^z$. 

However, for the non-integrable case shown in Figs.~(\ref{fig2},\ref{fig3}), it is clear that we do not get a convergence to the Mazur bound even on including higher powers of $H$. This is especially clear for the weak field case where it is expected that the system is highly non-ergodic and the non-convergence to the Mazur bound is a manifestation of this.

{\bf Quantum case}:We next consider the quantum case where the Hamiltonian consists of $1/2$ spins. We use the same parameter set $(\Delta ,h)=(0.5,0)$ for the integrable system, and the sets $(0.5,1.0)$ and $(0.5,0.1)$ for the non-integrable system. The set of conserved quantities that we use for integrable system is ${\bf I}_1=(1,S^z , H , S^z  H)$, while we consider three cases for non-integrable case, i.e., ${\bf I}_2=(1,H)$, ${\bf I}_3=(1,H,H^2)$, and ${\bf I}_4=(1,H,H^2,H^3)$. Then, we use the formula in Eq.~(\ref{mazur-suzuki}) to compute the corresponding Mazur bounds. We set $\beta=1.0$ in all cases.

\begin{figure}
  \centering
  \includegraphics[width=0.95\columnwidth]{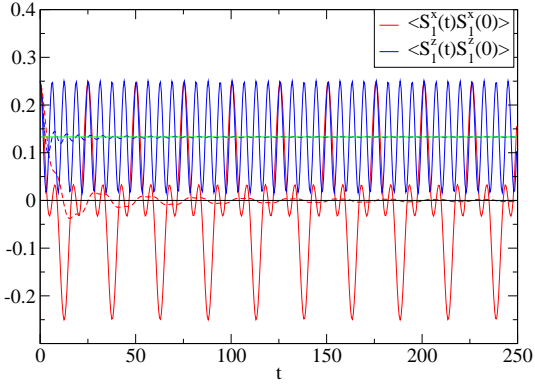}
  \caption{Quantum spins: Parameter values --- $\Delta = 0.5, h=0.0$. Plots of auto-correlation functions for $A=S_1^x$ and $A=S_1^z$, as obtained from direct simulations, compared with Mazur bounds for different choices of sets of conserved variables.  The dashed lines indicate integrals $t^{-1}\int_0^t dt'\la A(t')A(0)\ra$. The bounds shown at $D_A=0$ (black line) and $D_A=0.133106$ (green line) are obtained with the set ${\bf I}_1=\{1,S^z, H, S^z H \}$ and agree with the time average $C_A$.}
  \label{Qfig1}
\end{figure}

\begin{figure}
  \centering
  \includegraphics[width=0.95\columnwidth]{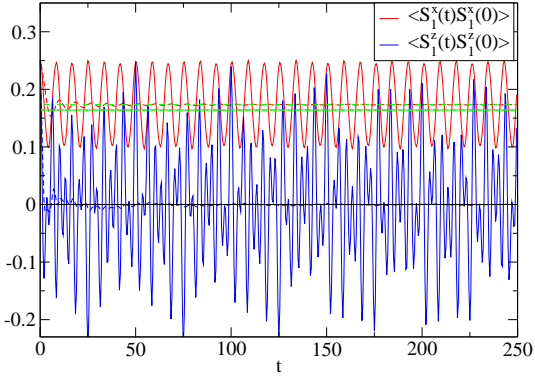}
  \caption{Quantum spins: Parameter values --- $\Delta = 1.5, h=1.0$. Plots of auto-correlation functions for $A=S_1^x$ and $A=S_1^z$, as obtained from direct simulations, compared with Mazur bounds for different choices of sets of conserved variables.  The dashed lines indicate integrals $t^{-1}\int_0^t dt'\la A(t')A(0)\ra$. 
  We see that the saturation values  quickly converge to the bound with a finite number of terms. For $A=S_1^x$, we get $C_A=0.173$ and the bounds $D_A^{(1)}=0.164074$ (green solid line), $D_A^{(2)}=0.164189$ (dashed green line) and $D_A^{(3)}=0.173$ (green dashed-dotted line), obtained respectively for the sets ${\bf I}=(1,H)$, ${\bf I}=(1,H,H^2)$ and ${\bf I}=(1,H,H^2,H^3)$.  For $A=S_1^z$, $D_A^{(n)}=0$ value (black line) is obtained for all $n$. }
  \label{Qfig2}
\end{figure}

\begin{figure}
  \centering
  \includegraphics[width=0.95\columnwidth]{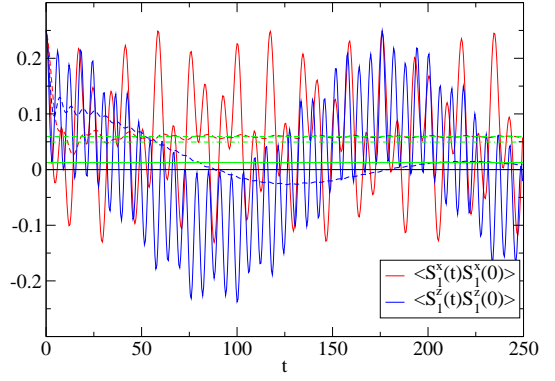}
  \caption{Quantum spins: Parameter values --- $\Delta = 1.5, h=0.1$. Plots of auto-correlation functions for $A=S_1^x$ and $A=S_1^z$, as obtained from direct simulations, compared with Mazur bounds for different choices of sets of conserved variables.  The dashed lines indicate integrals $t^{-1}\int_0^t dt'\la A(t')A(0)\ra$. 
For $A=S_1^x$, we get $C_A=0.0594$ and the bounds $D_A^{(1)}=0.0127$ (green solid line), $D_A^{(2)}=0.0495$ (dashed green line) and $D_A^{(3)}=0.0594$ (green dashed-dotted line), obtained respectively for the sets ${\bf I}=(1,H)$, ${\bf I}=(1,H,H^2)$ and ${\bf I}=(1,H,H^2,H^3)$.  For $A=S_1^z$, $D_A^{(n)}=0$ value (black line) is obtained for all $n$.
 We again see that the saturation values  quickly converge to the bound with a finite number of terms.}
  \label{Qfig3}
\end{figure}

In Figs.~(\ref{Qfig1},\ref{Qfig2},\ref{Qfig3}) we show the results for the three parameter sets. As expected for a finite quantum system with a discrete spectrum, the oscillations of the auto-correlation  do not die down, unlike the classical case. However, the long time average still exists of course, and we compare this with the Mazur bound.    

For the integrable case,  the numerical calculation of the exact long time limit for $A=S_1^z$ gives $C_A=0.133106$, while that for $A=S_1^x$ gives $C_A=0$. These values are perfectly reproduced by the formula (\ref{mazur-suzuki}) with the set ${\bf I}_1$. For the non-integrable case with $(\Delta,h)=(0.5,1.0)$, the exact value in the long time limit for $A=S_1^x$ is $0.173527$. For each set of conserved quantities, one can obtain the value by the formula (\ref{mazur-suzuki}); $0.164074$ for ${\bf I}_2$, $0.164189$ for ${\bf I}_3$, and $0.173527$ for ${\bf I}_4$. Thus we get the equality $C_A=D_A$ with a finite set of powers of $H$. For $A=S_1^z$, we get  $C_A=0$, and $D_A=0$ for all the choices of the set ${\bf I}$.
As seen in Fig.~\eqref{Qfig3} this is also seen for the parameter set $(\Delta ,h)=(0.5,0.1)$. The fact that we need up to $H^3$ for saturation of the Mazur bound, follows from the discussion around Eq.~\eqref{Hbasis}.

\subsection{Quadratic many-particle Hamiltonians}  
For the quantum case, we consider a fermionic system whose Hamiltonian can be expressed in the form 
\begin{align}
H= \sum_{i,j=1}^N [\gamma_{ij} a^\dagger_i a_j +\sum \Delta_{ij} (a^\dagger_i a^\dagger_j+ a_j a_i)] ~,  
\end{align}
and we take $\gamma$ and  $\Delta)$ to be  real symmetric  matrices. One can always consider a linear transformation to new fermionic variables $\{c_p\}$, $p=1,2,\ldots,N$, such that the Hamiltonian is reduced to the form $H=\sum_p \epsilon_p \hat{n}_p$, where $\hat{n}_p=c^\dagger_p c_p$ are then conserved quantities. Let us take $A=\sum_{pq} \alpha_{pq} c^\dagger_p c_q$ and let $\Delta A= A-\la A\ra$. 
Then we have 
\begin{align}
\la A(t) A(0) \ra &=  \sum_{p q} \sum_{r s} \alpha_{p q} \alpha_{r s} e^{i (\epsilon_p-\epsilon_q) t} \la c^\dagger_{p} c_{q} c^\dagger_r c_s \ra  \nn \\ 
&= \la A \ra ^2 +  \sum_{pq} {\alpha_{pq}^2} e^{i (\epsilon_p-\epsilon_q) t} \la c^\dagger_{p} c_{p}\ra \la  c_q c^\dagger_q \ra~, \nn \\
&=\la A \ra ^2 +  \sum_{pq} {\alpha_{pq}^2} e^{i (\epsilon_p-\epsilon_q) t} n_p (1-n_q)~, \\
{\rm where}~~\la A \ra&= \sum_k \alpha_{p p} n_p.
\end{align}
Performing a time average and assuming non-degenerate $\epsilon_p$ we get
\begin{align}  
C_A=\overline{\la \Delta A(t) \Delta A(0)\ra} = \sum_p \alpha_{pp}^2 n_p (1-n_p).
\label{DAferm} 
\end{align}
Next we compute the Mazur bound with the set ${\bf I}= (\Delta n_1, \Delta n_2,\ldots, \Delta n_N )$, where $\Delta n_p= n_p- \la n_p \ra$. We use the results
\begin{align}
\la (\Delta\hat{n}_p)^2 \ra&= n_p (1 -n_p) \nn \\
\la   \Delta A \Delta \hat{n}_p \ra &= \la A \hat{n}_p \ra - \la A\ra {n}_p \nn \\
&=\alpha_{pp} n_p (1-n_p)~, 
\end{align}
to get the Mazur bound
\begin{align}
D_A=\sum_p \f{ \la \Delta A \Delta \hat{n}_p \ra^2}{ \la  (\Delta\hat{n}_p)^2 \ra} = \sum_p  \alpha_{pp}^2 n_p (1-n_p)
\end{align}
and so we get equality to $C_A$ obtained in Eq.~\eqref{DAferm}.

For the harmonic crystal we consider the classical case though an extension to the quantum case is straight-forward. 
\begin{align}
H&= \sum_l \frac{p_l^2}{2} + \frac{1}{2} \sum_{l,m} \phi_{l,m} x_l x_m~ \\
&=\f{p^T p}{2}+ \frac{1}{2} x^T \phi x,
\end{align}
where  $\phi$ denotes the force matrix. Let $U$ be the normal mode transformation that is orthogonal $U^T U=I$ and diagonalizes $\phi$, i.e, $U^T \phi U^=\Omega^2$. Transforming  to normal mode coordinates $X=U^{-1} x$, where $X$ is a column vector $X_1,X_2,\ldots,X_N$ we get
\begin{align}
H&= \f{P^T P}{2}+\f{X^T \Omega^2 X}{2}= \sum_k H_k, \\
 {\rm where} ~~H_k &= \f{P_k^2}{2}+\Omega_k^2 \f{X_k^2}{2}.
\end{align}
For simplicity we will assume that the spectrum $\Omega_k$ is non-degenerate. 
For our observable we consider a quadratic form expressed in terms of the normal mode variables as $A(t)= \sum_{k,p} \alpha_{k p} X_k X_p$, where $\alpha_{k p}=\alpha_{p k}$ and then define $\Delta A(t)=A(t)-\la A \ra$.  Then we have 
\begin{align}
\la A(t)  A(0) \ra &=  \sum_{p q} \sum_{r s} \alpha_{p q} \alpha_{r s} \left\la \left[X_{p} \cos{\Omega_p t} +P_p \f{\sin {\Omega_p t}}{\Omega_p} \right] \nn \right. 
\\  &  \left. \times \left[ X_{q} \cos{\Omega_q t} +P_q \f{\sin {\Omega_q t}}{\Omega_q} \right] X_r X_s \right\ra \nn \\ 
&= \la A \ra ^2 + 2 T^2 \sum_{pq} \f{\alpha_{pq}^2}{\Omega_p^2 \Omega_q^2} \cos{\Omega_p t} \cos{\Omega_q t}~, \nn \\
{\rm where}~~\la A \ra&= T \sum_k \f{\alpha_{kk}}{\Omega_k^2}.
\end{align}
Hence, on performing a time average, we get
\begin{align}  
C_A=\overline{\la \Delta A(t) \Delta A(0)\ra} = \overline{\la A(t) A(0)\ra} -\la A \ra^2 =  T^2 \sum_p \f{\alpha_{pp}^2}{\Omega_p^4}. \label{harmtav}
\end{align}
For the Mazur bound we use the set ${\bf I}=(\Delta H_1,\Delta H_2,\ldots,\Delta H_N)$, where $\Delta H_p= H_p-\la H_p \ra$. We note that
\begin{align}
\la H_p\ra &= T,~\la H_p^2 \ra =2 T^2, \nn \\
\la \Delta H_p \Delta H_q \ra &= T^2 \delta_{pq} \nn \\
\la \Delta A \Delta H_p \ra &= \la A H_p \ra - \la A \ra \la H_p \ra = T^2 \f{\alpha_{pp}}{\Omega_p^2}.
\end{align}
Using these we compute  $\sum_{p} \f{\la  \Delta A \Delta H_p \ra^{2}}{\la (\Delta H_p)^2\ra}$ and find that this  precisely gives the expression in Eq.~\eqref{harmtav} and so we verify that the Mazur bound gives us the equality $C_A=D_A$. 

\subsection{Toda chain}
We next consider the many-body classical Toda chain which is an example of an interacting integrable system with non-trivial decay of current correlations~\cite{Spohn2018}. It is defined with the Hamiltonian~\cite{Toda1989a}:
\eqa{
	H(\{p_i,r_i\}) = \sum_{i=1}^N e_i = \sum_{i=1}^N \frac{p_i^2}{2} +  V(r_i),
}
where $r_i = x_{i+1}-x_i$ and $V(r) = e^{-r} $. We write the  equations of motion in the form
\begin{eqnarray}
  \dot{r_i} =& p_{i+1} - p_i,\\
  \dot{p_i} =& e^{-r_{i-1}} - e^{-r_i}, \label{eq:todaeom}
\end{eqnarray}
for $i=1,2,\ldots,N$, and  with the periodic boundary conditions $p_{N+1} = p_{1}, ~r_0 = r_N$.   These equations of motion can be cast in a Lax matrix form, namely,
$
	\frac{dL}{dt} = \lbrack M,L \rbrack \nonumber
$, where the matrix $L$ is defined as
\begin{align}
 L= \left( \begin{array}{cccc}
	b_1 & a_1  & ...&  a_N \\
	a_1 & b_2 & a_2 & 0  \\
	.&a_2&.&.\\
	.&.&.&.\\
	a_N & 0 & a_{N-1} & b_{N},  \\
	\end{array} \right),
\end{align}
with  $b_i =  p_i/2$ and $a_i = \frac{1}{2}e^{-r_i/2}$, 
while $M = L_{+} - L_{-}$ is the difference between the upper and lower triangular parts of matrix $L$.  
Since $M$ is an antisymmetric matrix, the eigenvalues of $L$ are time-independent  \cite{Flaschka1974, Henon} and the $N$ local independent conserved quantities
 $Q_{n}, n=1,2,\ldots,N$, of the Toda lattice can be written as
	\eqa{
Q_n = \frac{2} {(n-1)!} ~\text{Tr}~ [L^n]= \frac{2} {(n-1)!} \sum_{i=1}^{N}\left(L^{n}\right)_{i, i} .\label{eq:Todacons}
	}
In particular we see that  $Q_1 = \sum_i p_i$ is the total momentum, and $Q_2 = \sum_i e_i$ is the total energy of the system. Note that the above form 
means that the $n$-{th} conserved quantity  can be written as a sum over local  conserved densities $Q_n = \sum_i q_n(i)$, where $q_n(i)= [{2}/{(n-1)!}] \left(L^{n}\right)_{i, i}$ depends on the degrees of freedom of $n$ neighboring particles.  The explicit expressions of the first few conserved quantities up to an overall constant are given in \cite{young2010}.
	Apart from the  $N$ independent conservation laws, there is an extra conserved quantity $Q_0 = \sum_{i=1}^N r_i$, which is the total length of the periodic ring and is important for  the hydrodynamic description  of the system~\cite{Doyon2019a,Spohn2019b}.

Corresponding to the local conserved densities, $q_n(j)$ we can construct local currents. Using the Lax-matrix equation of motion one finds $\frac{d}{dt} q_n(i) =  \frac{2} {(n-1)!}\left(-2 a_i L^n_{i,i+1} + 2 a_{i-1} L^n_{i,i-1}\right)$  \cite{Spohn2019b,Spohn2019}, which has the form of a continuity equation $dq_n(i)/dt=j_n(i)-j_n(i+1)$ with the local current given by  $j_n(i) =  \frac{4} {(n-1)!} a_{i-1} L^n_{i,i-1}$. We define the corresponding total current $J_n = \sum_i j_n(i)$.
The first few local currents are the stretch current, $j_0(i)=-p_i$,   the momentum current, $j_1(i) = -V'(r_i) $,  and  the energy current, $j_2(i) = -p_i V'(r_i)$. In the thermodynamic limit, the average currents in the Generalized Gibbs ensemble in the Toda chain   have recently been computed  explicitly using the framework of generalized hydrodynamics~\cite{Spohn2019,Spohn2017}.

	Here we focus on the decay of  the correlations of the total momentum and energy currents, $J_1$ and $J_2$ in finite chains. Subtracting the mean we define 
\eqa{
A_n(t) = J_n(t) - \langle J_n \rangle ,
}
where the average $\la \ldots \ra$ is over the Gibbs ensemble  $\rho=e^{-\beta (H+P\sum_j r_j)}/Z(\beta,P)$, with $Z$ the normalization being specified by the inverse temperature $\beta$ and pressure $P>0$. We compute the quantity $\frac{1}{N}\la A_n(t)A_n(0)\ra , n=1,2$ and its long time average from microscopic evolution of Eq.~\ref{eq:todaeom}. At large times the correlation is bounded by the Mazur value given by Eq.~\eqref{mazur-suzuki}. We compute this bound numerically, for different choices of the set $\{I_s\}$, as an equilibrium average over the same Gibbs ensemble and compare with the corresponding time average. 

The decay of the energy current in this model was studied in \cite{young2010}, where it was found that for finite chains ($N=4,6,8$), it was in fact necessary to consider projections  of the energy current to products of the independent conserved quantities, in order  to obtain saturation of the Mazur bound. Here, using the ideas developed in \cite{Spohn2019} we use a numerical approach that uses the Lax-matrix construction to compute the equal time correlations. The basic idea is to first note that the Gibbs measure is in the product form $\rho=\prod_{i=1}^N e^{-\beta (p_i^2/2+V(r_i)+P r_i)}/Z$, with $Z=\int_{-\infty}^\infty dp \int_{-\infty}^\infty dr e^{-\beta (p^2/2+V(r)+P r)}$, which means that the elements of the Lax-matrix,  $\{ b_i \}$ are Gaussian distributed, while the  $\{a_i \}$ are from a Chi-square distribution \cite{Spohn2019}. Secondly, we have seen that the currents are expressible in terms of the Lax-matrix. Hence an efficient numerical scheme is to generate an an ensemble of $L$ matrix copies and find the required averages. Using this approach we are able to study systems up to size $N=10$.
In our studies we set $\beta=1,P=1$ and the averages, for both sides of  Eq.~\eqref{mazur-suzuki}, are obtained over  $\approx 10^8-10^{9}$ samples. 
\begin{figure}
  \centering
  \includegraphics[width=\linewidth]{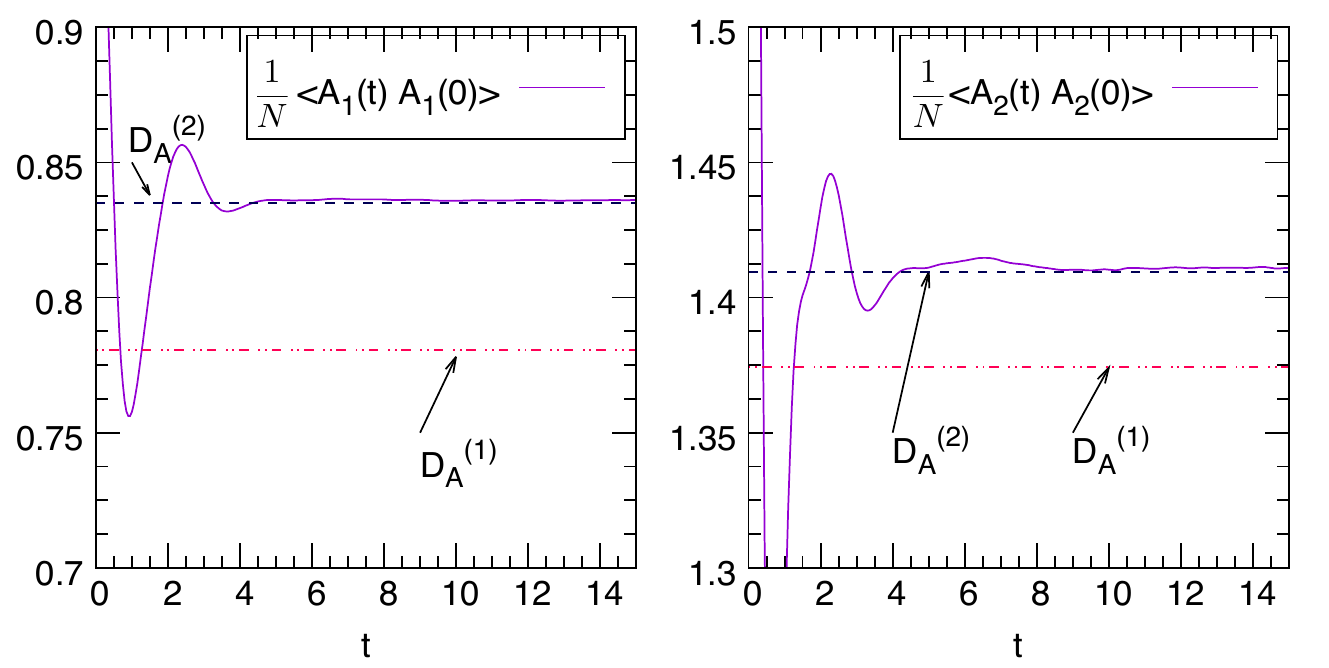}
  \caption{
    Toda chain with $N=4$: Plots of auto-correlation functions for momentum current $\langle A_1(t) A_1(0) \rangle$ and energy current $\langle A_2(t) A_2(0) \rangle$  (solid lines).  The  Mazur bounds for momentum current $D_A^{(1)}=0.7804$ from the set  ${\bf I}_1$ and  $D_A^{(2)}=0.8379$ from the set ${\bf I}_2$ are shown as dashed lines.  For the energy current, the  corresponding Mazur bounds are $D_A^{(1)}=1.3741$ and  $D_A^{(2)}=1.4094$. The temperature and pressure of the Gibbs ensemble were taken as $\beta=P=1$.}
  \label{fig:N4toda}
\end{figure}

\begin{figure}
  \centering
  \includegraphics[width=\linewidth]{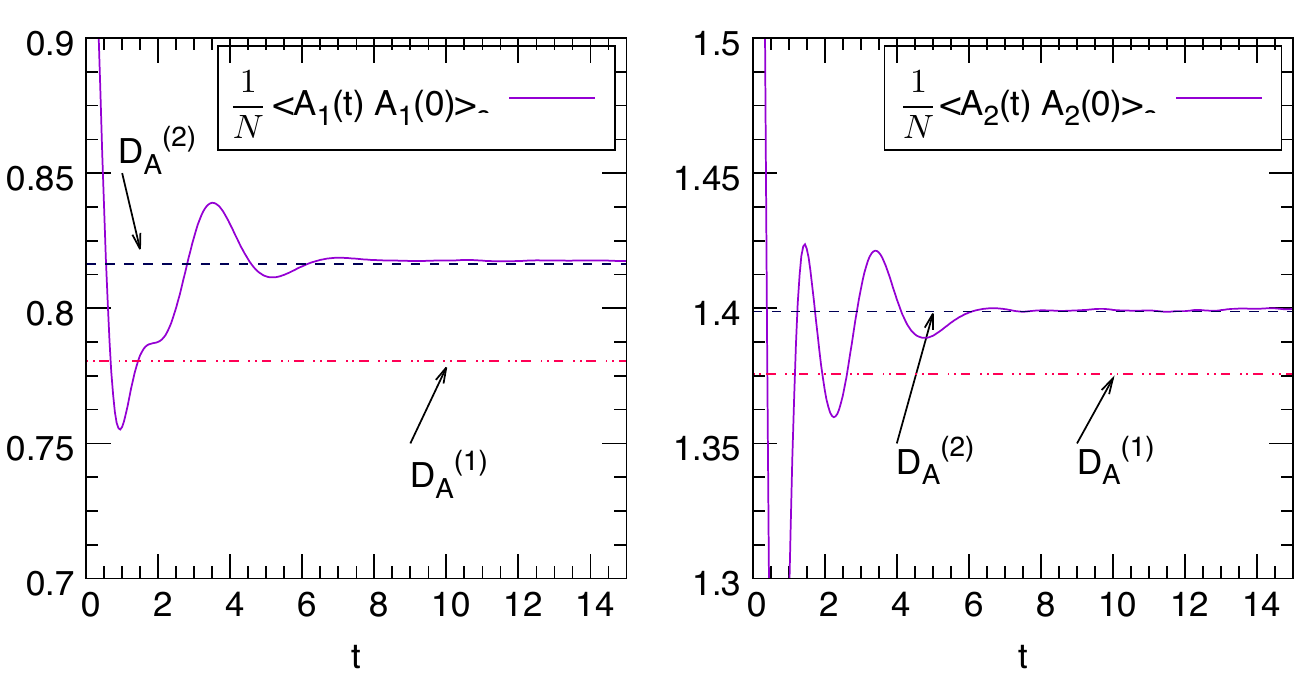}
  \caption{ Toda chain with $N=6$: Plots of auto-correlation functions for momentum current $\langle A_1(t) A_1(0) \rangle$ and energy current $\langle A_2(t) A_2(0) \rangle$  (solid lines).  The  Mazur bounds for momentum current $D_A^{(1)}=0.7804$ from the set  ${\bf I}_1$ and  $D_A^{(2)}=0.8163$ from the set ${\bf I}_2$ are shown as dashed lines.  For the energy current, the  corresponding Mazur bounds are $D_A^{(1)}=1.3757$ and  $D_A^{(2)}=1.3981$. All parameters are as in Fig.~\ref{fig:N4toda}.}
  \label{fig:N6toda}
\end{figure}
\begin{figure}
	\centering
	\includegraphics[width=\linewidth]{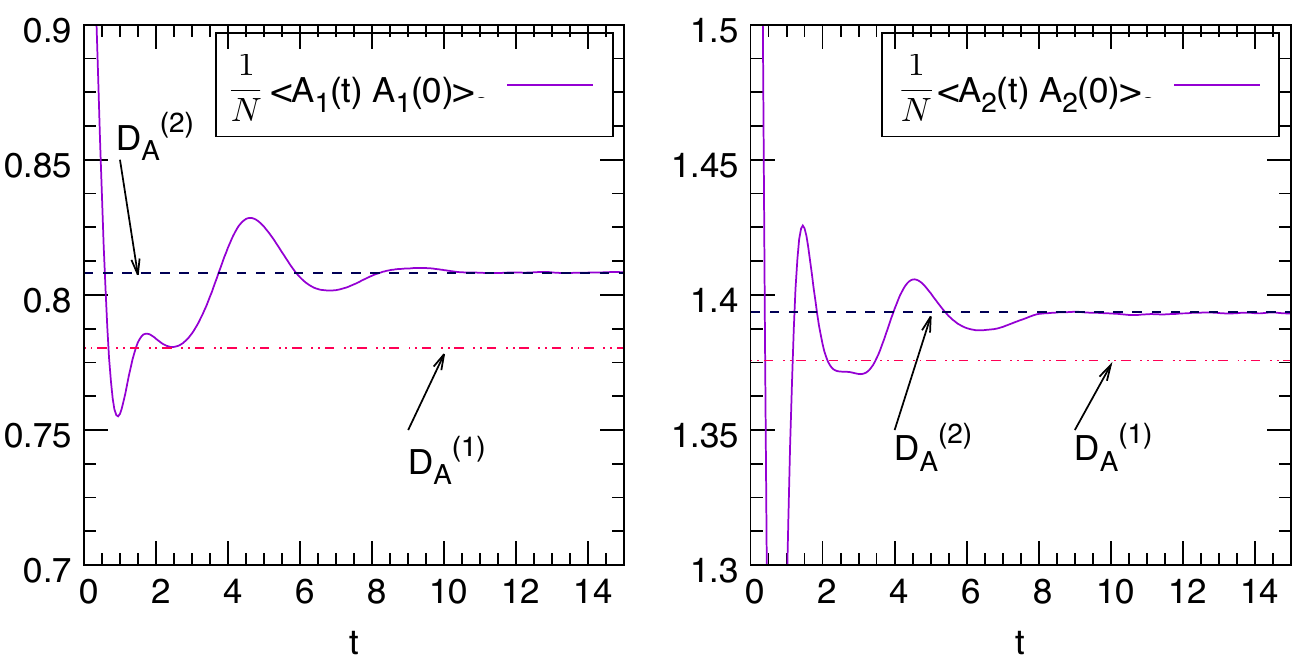}
	\caption{ Toda chain with $N=8$: Plots of auto-correlation functions for momentum current $\langle A_1(t) A_1(0) \rangle$ and energy current $\langle A_2(t) A_2(0) \rangle$  (solid lines).  The  Mazur bounds for momentum current $D_A^{(1)}=0.7804$ from the set  ${\bf I}_1$ and  $D_A^{(2)}=0.8083$ from the set ${\bf I}_2$ are shown as dashed lines.  For the energy current, the  corresponding Mazur bounds are $D_A^{(1)}=1.3757$ and  $D_A^{(2)}=1.3932$. All parameters are as in Fig.~\ref{fig:N4toda}.}
	\label{fig:N8toda}
\end{figure}

\begin{figure}
\centering
\includegraphics[width=\linewidth]{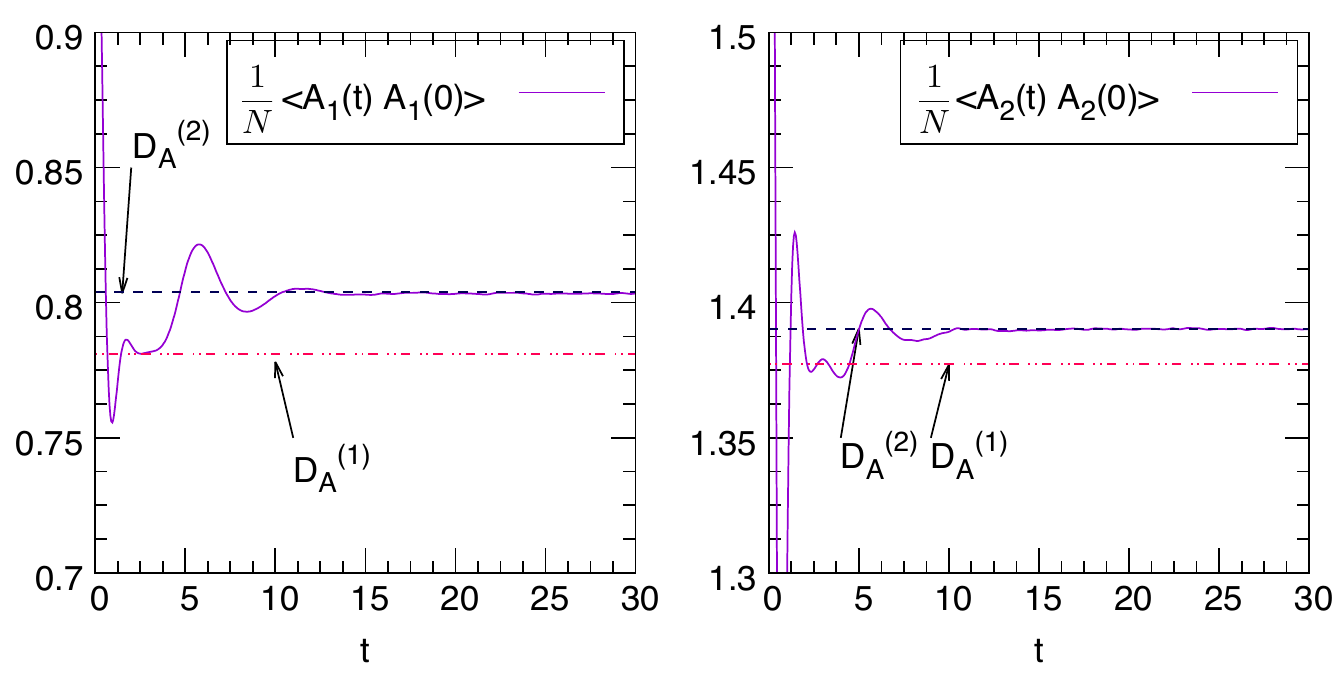}
\caption{Toda chain with $N=10$: Plots of auto-correlation functions for momentum current $\langle A_1(t) A_1(0) \rangle$ and energy current $\langle A_2(t) A_2(0) \rangle$  (solid lines).  The  Mazur bounds for momentum current $D_A^{(1)}=0.7811$ from the set  ${\bf I}_1$ and  $D_A^{(2)}=0.8036$ from the set ${\bf I}_2$ are shown as dashed lines.  For the energy current, the  corresponding Mazur bounds are $D_A^{(1)}=1.3772$ and  $D_A^{(2)}=1.3905$. All parameters are as in Fig.~\ref{fig:N4toda}.}
	\label{fig:N10toda}
\end{figure}

	We test Mazur bounds for the momentum current ($A_1$)  and the energy current  ($A_2$) of the Toda chain in Fig. \eqref{fig:N4toda} for $N=4$ and for $N=6$ in Fig. \eqref{fig:N6toda}. The long time decay of current auto-correlation  is compared with Mazur bounds for two different choices of sets of conserved variables: the first set involves $N+1$  conserved quantities (${\bf I}_1 = \{Q_n\},~ n=0 \dots N$), while the second set has  the $N+1$ conserved quantities along with their $(N+1)(N+2)/2$ products (${\bf I}_2=\{Q_n, Q_n Q_m\}, ~n=0 \dots N$).  Note that  $A_1$ is even under time reversal symmetry which makes its non-zero overlap only with  even conserved charges (i.e. $0,2,4,....$), while $A_2$ is odd under time reversal which has non-zero overlap with odd conserved charges or products of conservation charges which are odd. 

In Figs.~(\ref{fig:N4toda},\ref{fig:N6toda},\ref{fig:N8toda},\ref{fig:N10toda}), we plot the auto-correlation functions of $A_1$ and $A_2$ and compare them with the Mazur bounds obtained with the two conserved sets ${\bf I}_1$ and ${\bf I}_2$. The time-averaged correlation value, $C_A$ and the Mazur bounds are tabulated in Tables~(\ref{tab:diffmomentum},\ref{tab:diffenergy}).
\begin{table*}
  {\centering
        \begin{tabular}{lllll}
  	
  	N                        &~~$D_A^{(1)}$~~  & ~~$D_A^{(2)}$   ~~              &  ~~$C_A$              & $C_A- D_A^{(1)}$                  \\ \hline
  	\multicolumn{1}{|l|}{4}  & \multicolumn{1}{l|}{0.7805 $\pm$ 0.0003} & \multicolumn{1}{l|}{0.8349  $\pm$  0.0003} &\multicolumn{1}{l|}{0.8360  $\pm$  0.00016} & \multicolumn{1}{l|}{0.056} \\ \hline
  	\multicolumn{1}{|l|}{6}  & \multicolumn{1}{l|}{0.7804 $\pm$ 0.0003} &\multicolumn{1}{l|}{0.8165  $\pm$  0.0003} & \multicolumn{1}{l|}{0.8176  $\pm$  0.00016} & \multicolumn{1}{l|}{0.037} \\ \hline
  	\multicolumn{1}{|l|}{8}  & \multicolumn{1}{l|}{0.7802 $\pm$ 0.0003} &\multicolumn{1}{l|}{0.8083 $\pm$  0.0003} & \multicolumn{1}{l|}{0.8083 $\pm$ 0.00018} & \multicolumn{1}{l|}{0.028} \\ \hline
  	\multicolumn{1}{|l|}{10}  & \multicolumn{1}{l|}{0.7811 $\pm$ 0.0003} &\multicolumn{1}{l|}{0.8036 $\pm$ 0.0003} & \multicolumn{1}{l|}{0.8035 $\pm$ 0.00017} & \multicolumn{1}{l|}{0.022} \\ \hline			
  \end{tabular} 
  }
\caption{Values of the time-averaged momentum current auto-correlation, $C_A = \frac{1}{T}\int_0^T \la A_1(t) A_1(0)\ra dt$, and the Mazur values $D_A^{(1)}$ and  $D_A^{(2)}$ for increasing system sizes.}
\label{tab:diffmomentum}
\end{table*}

\begin{table*}
	\centering
	\begin{tabular}{lllll}
	N                        &~~$D_A^{(1)}$~~  & ~~$D_A^{(2)}$   ~~              &  ~~$C_A$              & $C_A- D_A^{(1)}$                  \\ \hline
	\multicolumn{1}{|l|}{4}  & \multicolumn{1}{l|}{1.3742 $\pm$ 0.0005} & \multicolumn{1}{l|}{1.4094 $\pm$ 0.0007} &\multicolumn{1}{l|}{1.4109 $\pm$ 0.0003} & \multicolumn{1}{l|}{0.037} \\ \hline
	\multicolumn{1}{|l|}{6}  & \multicolumn{1}{l|}{1.3757 $\pm$ 0.0005} &\multicolumn{1}{l|}{1.3988 $\pm$ 0.0007} & \multicolumn{1}{l|}{1.3993 $\pm$ 0.0003} & \multicolumn{1}{l|}{0.023} \\ \hline
	\multicolumn{1}{|l|}{8}  & \multicolumn{1}{l|}{1.3766 $\pm$ 0.0005} &\multicolumn{1}{l|}{1.3937 $\pm$ 0.0002} & \multicolumn{1}{l|}{1.3933  $\pm$ 0.0003} & \multicolumn{1}{l|}{0.017} \\ \hline
	
	\multicolumn{1}{|l|}{10}  & \multicolumn{1}{l|}{1.3770 $\pm$ 0.0005} &\multicolumn{1}{l|}{1.3905 $\pm$ 0.0003} & \multicolumn{1}{l|}{1.3901 $\pm$ 0.0003} & \multicolumn{1}{l|}{0.013} \\ \hline
\end{tabular} 
	\caption{Values of the time-averaged energy current auto-correlation, $C_A = \frac{1}{T}\int_0^T \la A_2(t) A_2(0)\ra dt$, and the Mazur values $D_A^{(1)}$ and  $D_A^{(2)}$ for increasing system sizes.} \label{tab:diffenergy}
\end{table*}

\begin{figure}
	\centering
	\includegraphics[width=\linewidth]{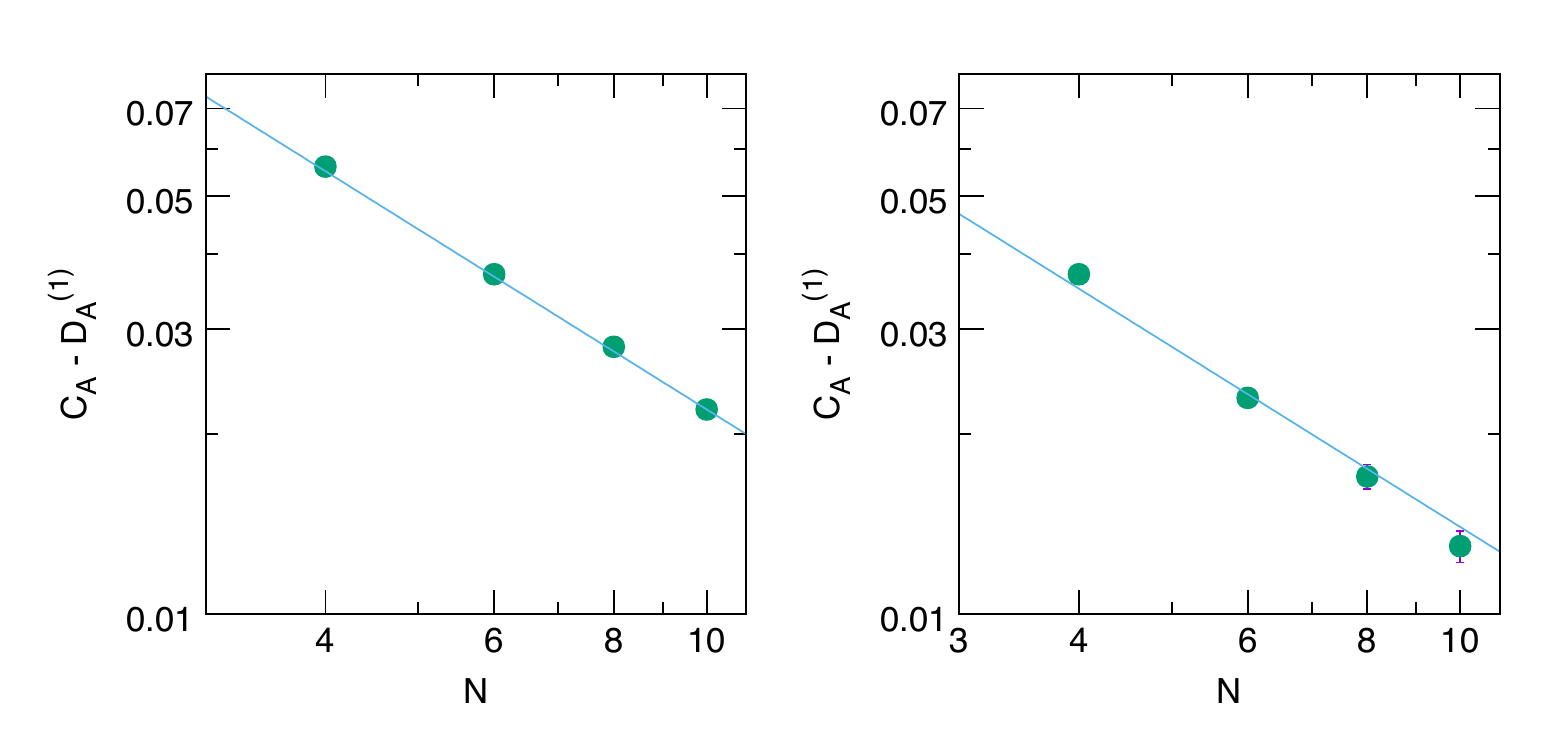}
	\caption{Toda chain: The difference of  $C_A - D_A^{(1)}$,  plotted in log-log scale, as a function of $N$ for the momentum current (left panel) and the energy current of the Toda chain. The data is enumerated in Tables~[\ref{tab:diffmomentum}] and [\ref{tab:diffenergy}]. The solid lines have slopes $-1$ and indicate a $\sim 1/N$ decay.  }
	\label{fig:differencendependence}
\end{figure}

{\bf Main observation}: Consistent with the observations in \cite{young2010} we see here that, for both the momentum and energy currents, for finite chains the  equality $C_A=D_A^{(2)}$ seems to be obtained (up to our numerical error bars), where $D_A^{(2)}$ is constructed from the set ${\bf I}_2$ that includes both the independent conserved quantities $\{Q_n\}$, as well as all products $\{Q_n Q_j\}$. Secondly as seen from the entries in the fifth column of Tables~(\ref{tab:diffmomentum},\ref{tab:diffenergy}), we  see that with increasing chain length the relative difference $(C_A-D_A^{(1)})/C_A$ seems to be decreasing. In  Fig.~\ref{fig:differencendependence} we see that the absolute difference $C_A - D_A^{(1)}$ decreases as $1/N$ with increasing system size.   It is thus plausible that  the equality  $C_A=D_A^{(1)}$ is obtained in the thermodynamic limit.

\section{Discussion}
\label{sec:summary}
We examined the relation between the Mazur inequality and the Suzuki equality. 
In particular we asked as to when, for classical systems, the Mazur inequality become an equality. In that case the time averaged autocorrelation, $C_A$, of an observable $A$ would be exactly equal to the Mazur bound, $D_A$.  
A crucial point is the choice of conserved quantities to be included while constructing the $D_A$. In general, a classical system with $N$ coordinate degrees of freedom will have a small number of independent conserved quantities, while integrable systems have exactly $N$ independent conserved quantities. We label these independent conserved quantities as ${\bf Q}$. Then we argue that an equality between $C_A$ and $D_A$ can be obtained if the dynamics is ergodic in the restricted phase-space (microcanonical surface with fixed constants of motion). However, for a finite system, while constructing $D_A$ it is not sufficient to consider only the set ${\bf Q}$ but also in general, all  higher powers such as $\{Q_j Q_k, Q_j Q_k Q_l,\ldots \}$.  The set of conserved quantities used in constructing $D_A$ was denoted by ${\bf I}$. The numerical examples of an anharmonic oscillator and two coupled spins were used to illustrate these points. The coupled spin system is on example where the equality is not satisfied, presumably because of lack of ergodicity.  

The set ${\bf I}$ required to get the equality of course  depends on the  observable. We showed that for quadratic Hamiltonians, either classical or quantum, for a quadratic observable, the set ${\bf Q}$ was sufficient while for a quartic observable, the set $\{Q_j Q_k\}$ has to be added.  
We note that  related ideas have been discussed in the context of eigenstate hypothesis (ETH), where it has been pointed out that in order to quantify diagonal fluctuations, it is necessary to take not just projections on local conserved 
quantities but also  products of conserved quantities~\cite{Vidmar2020}. Another intriguing idea that has been proposed is the idea of quasilocal constants of motion which, via the Mazur inequality, lead to  rigorous bounds on the spin Drude weight in integrable quantum spin chains \cite{prosen2011} and possibly in a related classical model \cite{prosenLLL2013}.

Finally we considered the integrable Toda chain and numerically studied the momentum and energy current correlations and found that, in order to get the Mazur equality, one needs to consider not just the $N$ independent constants of motion, but also higher powers. We presented evidence that the contribution of higher powers  decrease with system size as $1/N$ and could vanish in the thermodynamic limit. We are not aware of a proof of this, though one can imagine one along the lines leading to Eq.~\eqref{ergodic}.   While this seems straight-forward for a system with a finite number of constants of motion, this may be less trivial to demonstrate for classical integrable system with a macroscopic number of constants of motion. 

\section{Acknowledgement}
We thank Sriram Shastry and Peter Young for very useful comments and suggestions.
A.D. acknowledges support of the Department of Atomic Energy, Government of India, under project no.12-R\& D-TFR-5.10-1100.
K.S. was supported by Grants-in-Aid for Scientific Research (JP16H02211, JP19H05603, JP19H05791).

\end{document}